\documentstyle[12pt]{article}
\newcommand{\bq}{\begin{equation}}
\newcommand{\ee}{\end{equation}}
\newcommand{\bi}[1]{\bibitem{#1}}
\newcommand{\fr}[2]{\frac{#1}{#2}}
\newcommand{\bqas}{\begin{eqnarray*}}
\newcommand{\eeas}{\end{eqnarray*}}
\newcommand{\non}{\nonumber}
\newcommand{\bqa}{\begin{eqnarray}}
\newcommand{\eea}{\end{eqnarray}}

\begin{document}
\pagestyle{plain}
\pagenumbering{arabic}
\vspace{1.0cm}
\begin{center}{\Large \bf  Instanton -- Antiinstanton interaction and
asymptotics of perturbation theory expansion in double well oscillator }\\
\vspace{1.0cm}

{\bf  S.V.Faleev, P.G.Silvestrov}\footnote{e-mail address:
PSILVESTROV@INP.NSK.SU} \\
Budker Institute of Nuclear Physics and Novosibirsk University, \\ 630090
Novosibirsk, Russia

\vspace{0.5cm}

March 10,1994

\vspace{1.5cm}

\end{center}

\begin{flushright}
{\bf BUDKERINP 94--24}
\end{flushright}

\vspace{1.0cm}

\begin{abstract}

Instanton -- antiinstanton pair is considered as a source of singularity
at the Borel plane for the ground state energy of anharmonic oscillator. The
problem of defining the short range instanton -- antiinstanton interaction
reduces to calculation of a smooth part of the Borel function, which cannot
be found without explicit calculation of several terms of ordinary
perturbation theory. On the other hand, the large order terms of perturbative
expansion are dominated by large fluctuations in the functional integral like
well separated instanton and antiinstanton.

The preasymptotics ($\sim 1/n$) of large order perturbation theory
contribution to the ground state energy of anharmonic oscillator was found
analytically. To this end the subleading long range asymptotics of the
classical instanton -- antiinstanton interaction, the one -- loop quantum
contribution to instanton -- antiinstanton interaction and the second
quantum correction to a single instanton density were considered.

\end{abstract}

\newpage

\section{Introduction}\label{sec:1}

In this note we would like to consider how the large fluctuations in the
Euclidean functional integral contribute to the ground state energy of double
well anharmonic oscillator
\bq\label{eq:Ham}
H = - \fr{1}{2}\fr{d^2}{d x^2}
+\fr{1}{2} x^2 (1-gx)^2 \,\,\,\, .
\ee
Our practical aim will be the analytical derivation of the $\sim 1/n$
correction to the instanton -- antiinstanton pair induced asymptotics of the
perturbative expansion:
\bqa\label{eq:Asip}
E=\sum E_n g^{2n} \,\, ,
E_n=-\fr{3^{n+1}n!}{\pi}(1-\fr{53}{18n}+\ldots)\,\,.
\eea

After the works of L.N.Lipatov \cite{Lipatov} it is generally recognized,
that classical paths in the Feynman functional integral contribute
significantly to the asymptotics of the ordinary perturbation theory. However
the attraction between instanton and antiinstanton makes the high order
estimates for the double well oscillator (\ref{eq:Ham}) somewhat more
complicated. The usual trick to avoid this problem is the analytical
continuation of the instanton -- antiinstanton contribution over the
coupling constant \cite{Bogomol}.

In present paper we prefer not to make the ambiguous analytical continuation
at the intermediate steps of approximate calculation. Following t'Hooft
\cite{t'Hooft} (see also \cite{Balitsky}) we transform the functional
integral to the Borel type integral by considering the action as a
collective variable. In this approach the instanton -- antiinstanton pair
manifests itself as the singularity of the Borel transform. The problem of
short range instanton -- antiinstanton interaction reduces to that of
calculation the shape of Borel transform far from singular point. On the
other hand the smooth part of the Borel function is sensitive mostly to a
first few terms of perturbative expansion. Thus not only instanton helps one
to find the asymptotics of perturbation theory, but also the precise
calculation of  several terms of perturbative expansion allows to avoid the
problem of short range instanton -- antiinstanton attraction.

The type of singularity at the Borel plane is determined by interaction
between the instanton and antiinstanton. Therefore one have to go beyond the
dilute gas approximation in order to find the asymptotics of perturbative
expansion. Up to now all authors have taken into account only the leading
$\sim e^{-T}/g^2$ term of the classical instanton -- antiinstanton
interaction ($T$ is the distance between the pseudoparticles). In order to
find the $\sim 1/n$ correction to (\ref{eq:Asip}) we have to consider the
subleading $\sim T e^{-2T}/g^2$ correction to classical instanton --
antiinstanton interaction, the one loop quantum correction $\sim T e^{-T}$
to the interaction and the second quantum correction $\sim g^2$ to the
single instanton density.

\section{Asymptotics of the perturbative expansion and nonperturbative
effects}
\label{sec:2}

The ground state energy for the Hamiltonian (\ref{eq:Ham}) is given by the
functional integral
\bqa\label{eq:S}
E(g)=\lim_{L\to\infty}-\fr{1}{L}\ln Tr e^{-HL}\,\, &,& \,\,
 Tre^{-HL}=N\int Dx(t)\, e^{-S} \,\, ,
 \nonumber \\
S = \int_{-\fr{L}{2}}^\fr{L}{2}dt \biggl( \fr{1}{2}\dot x^2&+&
 \fr{1}{2}x^2(1-gx)^2\biggr)\,\,.
\eea
Instanton -- the well known classical path reads:
\bq\label{eq:Inst}
x_I(t)=\fr{1}{g(e^t+1)} \quad , \quad S[x_I]=\fr{1}{6g^2}\,\,\, .
\ee
This path connects two distinct vacua of the action (\ref{eq:S})
$x=0$ and $ x=1/g $. We want to find the configurations, which make a
large contribution to the high order terms of perturbation theory and
therefore may be considered as a fluctuation (although large) around the
trivial vacuum $x=0$. The example of such topologically trivial fluctuation
is the instanton -- antiinstanton pair. The pair is not the exact solution
of the equation of motion and therefore we have to understand, what the
specific configuration may be called the instanton -- antiinstanton pair?
But to a first approximation the interaction between pseudoparticles is model
independent for a wide class of reasonable configurations (see e.g.
\cite{Zinn-Justin})
\bq\label{eq:SInst}
 S[x_{I-A}]=\fr{1}{g^2}(1/3-2e^{-T}+\cdots) \,\,\, .
\ee
The instanton -- antiinstanton contribution to $ Tr e^{-HL} $ is given by:
\bq\label{eq:tr2}
(Tr e^{-HL})^{I-A}\simeq L \int \fr{dT}{\pi g^2}
exp\{-\fr{1}{g^2}(1/3-2e^{-T}+\cdots)\} \,\,.
\ee
Factor $L$ appears after integration over the center of the pair and $ (\pi
g^2)^{-1}$ comes from the square of single instanton density (see \cite
{Coulmen}). In order to find the ground state energy one have to sum over
the "dilute gas of instanton -- antiinstanton pairs". Hence the energy
reads:
\bq\label{eq:En}
E(g)=-\int \fr{dT}{\pi g^2}exp\{-\fr{1}{g^2}(1/3-2e^{-T})\}.
\ee
This integral diverges at large $T$. The divergency will disappear if for
large $T$ one replaces the summation over dilute gas of pairs by summation
over dilute gas of individual pseudoparticles. Nevertheless even the
expression (\ref{eq:En}) carries quite sufficient information about high
orders of perturbation theory. Consider new variable $x=g^2 S$, where $S$ --
is a classical action of the instanton -- antiinstanton pair. The value of
$x$ varies from $x=0$ (the classical vacuum) to $x=1/3$ and for widely
separated pseudoparticles $x=1/3-2e^{-T}+...$ . Now formula ({\ref{eq:En})
takes the form:
\bq\label{eq:Fbor}
E=\int_{0}^{1/3} F(x) e^{-x/g^2} \fr{dx}{\pi g^2} \,\, ,\,\,
 F(x)\approx -\fr{1}{(1/3-x)} (1+O(g^2)) + \ldots \,\, .
\ee

About the function $F(x)$ we know only that it has a pole at $x=1/3$.
However just the singularities of $F(x)$ at complex $x$ plane determine the
behaviour of series $F(x)=\sum F_n x^n$ at large $n$ and thus the
asymptotics of perturbative expansion of $E(g)$. The only assumption that
pole  at $x=1/3$ (\ref{eq:Fbor}) is the closest to origin singularity of
$F(x)$ allows one to find the well known \cite{Brezin} leading term of the
asymptotic expansion (\ref{eq:Asip}). Below we will find the weaker
(logarithmic) singularity of function $F(x)$ at the same point $x=1/3$.
There is no direct way to calculate the smooth function $F(x)$ at $x<1/3$,
but after one have calculated precisely several terms of $E(g)$ perturbative
expansion, the smooth part of $F(x)$ carries now new information.

At very large $T$ the dilute gas of many pseudoparticles is to be
considered. In order to eliminate the double counting one have to subtract
from (\ref{eq:En}) the contribution of noninteracting instanton and
antiinstanton:
\bq\label{eq:En1}
E_{\pm}=
-\biggl(\int_{S(T)=0}^{T=\infty}\fr{dT}{\pi g^2} e^{-S(T)}
-\int_{T=0}^{T=\infty}\fr{dT}{\pi g^2} e^{-1/3g^2}\biggr)
\mp \fr{e^{-1/6g^2}}{g\sqrt{\pi}}\,\,.
\ee
Here the last term describes the nonperturbative splitting of the ground
state due to tunneling through the barrier. Two signs correspond to the
states having different parity with respect to transformation $x \rightarrow
1/g-x$. After passing to variable $x=g^2 S$ the eq. (\ref{eq:En1}) takes the
form:
\bq\label{eq:En3}
E_{\pm}=-\biggl(
\int_{0}^{1/3}\fr{e^{-x/g^2}-e^{-1/3g^2}}{1/3-x}\fr{dx}{\pi g^2}
 -\fr{e^{-1/3g^2}\ln(6)}{\pi g^2}
\biggr)
\mp \fr{e^{-1/6g^2}}{g\sqrt{\pi}} \,\,.
\ee
This expression correctly accounts for the high order terms of perturbation
theory and the main nonperturbative effects caused by single instanton or
instanton -- antiinstanton pair. However the first terms of perturbative
expansion are completely wrong in (\ref{eq:En3}). In order to make
(\ref{eq:En3}) quantitatively reliable one have to calculate precisely the
$N$ first terms of the perturbation theory and subtract from (\ref{eq:En3})
the $N$ first terms of its expansion in series in $g^2$:
\bqa\label{eq:Etoch}
E_{\pm}&=&\sum_{n=0}^Ng^{2n}E_n
-\biggl\{ \int_{0}^{1/3}
\fr{e^{-x/g^2}-e^{-1/3g^2}}{1/3-x}\fr{dx}{\pi g^2}
\!\,-\fr{e^{-1/3g^2}ln(6)}{\pi g^2} -
\! \sum_{n=0}^N \fr{3^{n+1} (n)!}{\pi} g^{2n}\biggr\} \non \\
& & \mp \fr{e^{-1/6g^2}}{g\sqrt{\pi}}(1+\sum_{n=1} g^{2n}E_n^1)\,\,.
\eea
Here the last term $\sim\exp(-1/6g^2)$ formally is the largest
nonperturbative correction, but this term cancels if one considers the
"center of band" $E_{cb}= \fr{1}{2}(E_+ + E_-)$. The terms of series
(\ref{eq:Asip}) begin to grow up at $n>1/3g^2$. Respectively the approximate
expression (\ref{eq:Etoch}) reaches the best accuracy at $N=1/3g^2$.
Calculating the new terms of perturbation theory at $N>1/3g^2$ one should
only decrease the accuracy of total results.

In addition to the "center of band" the formula (\ref{eq:Etoch}) allows to
calculate also the individual energy levels. One have to calculate only a
lot of corrections $\sim g^{2n}$ to the instanton induced splitting of the
ground state energy $\sim e^{-1/6g^2}/g$. The asymptotics of perturbative
expansion of energy shift takes the form (see \cite{Zinn-Justin Mat}):
\begin{eqnarray}\label{eq:DE}
\Delta E=
\fr{2e^{-1/6g^2}}{g\sqrt{\pi}}(1+\sum_{n=1} g^{2n}E_n^1)
\,\,\, , \,\,\,
E_n^1=-\fr{3^{n+2}n!}{\pi}(ln(6n)+\gamma)\,\, ,
\end{eqnarray}
where $\gamma=0.5772 ...$ is the Eurlers constant. The perturbative
series for the energy shift looks very similar with that for the ground
state energy. The "best accuracy" one can reach after summing this series
(which is not worse than the smallest term of the series) is negligibly small
$ (\sim \exp(-1/2g^2)) $. In order to reach the accuracy $\sim
\exp(-1/3g^2)/g^2$  in $E_+$ or $E_-$ one have to break the summation in
(\ref{eq:DE}) at the term for which $g^{2n}E^1_n  \ll  \exp(-1/6g^2)$. The
three -- instanton contribution to $\Delta E$, which generates the asymptotics
(\ref{eq:DE}) is considered in the Appendix (see (\ref{eq:DeltaE})).

\section{ Corrections to effective action}
\label{sec:3}

Up to now we have taken into account only the leading term of interaction
between instanton and antiinstanton $\Delta S \sim e^{-T}/g^2$
(\ref{eq:SInst}). The further correction to the action of the pair will be
considered in this section.

The corrections of two types to the effective interaction are to be taken
into account. The first is $\Delta S_c \sim e^{-2T}/g^2$ -- the correction
to the classical instanton -- antiinstanton interaction. The second
correction $\Delta S_G \sim e^{-T}$ appears after the Gaussian integration
around the classical configuration. At first sight it seems that both these
corrections do not contribute to the asymptotics of the perturbation theory.
For example let us substitute the one loop correction $\Delta S_G \sim
e^{-T}$ in the equation (\ref{eq:En}):
\bq\label{eq:dEn}
E(g)=-\int \fr{dT}{\pi g^2} \exp\left\{ -\fr{1}{g^2}(1/3-2e^{-T})
\right\} \left( 1+ B e^{-T}\right) \,\, .
\ee
Here $B\sim 1$. Passing to the variable $x=g^2 S$ we get:
\bq\label{eq:dEn1}
E(g)= -\int \fr{e^{-x/g^2}}{1/3-x}  \left[ 1+\fr{1}{2}
B(1/3-x) \right]\fr{dx}{\pi g^2} \,\, .
\ee
The term proportional to $B$ cancels in the singularity of the
Borel function $F(x)$ (\ref{eq:Fbor}) and therefore does not contribute to
the asymptotics of the perturbation theory. It is easy to show that
correction to the classical action $\Delta S_c \sim e^{-2T}/g^2$ also does
not contribute to the asymptotics.

Nevertheless as we will see below the corrections to the Gaussian integral
and to the classical interaction appear which contain an additional large
factor $T$: $\Delta S_G \sim T e^{-T}$ and $\Delta S_c \sim T e^{-2T}/g^2$.
Taking into account of these corrections leads to a new singularity of the
Borel function $\delta F\sim\log(1/3-x)$, although weaker than the pole
(\ref{eq:Fbor}).

The quantum correction to the two instanton density is also to be taken into
account:
\bq\label{eq:rhoIA}
\rho = \rho_I \rho_A = \fr{1}{\pi g^2} \exp(-2S_I)(1+A g^2) \,\, .
\ee
Because of the correction $Ag^2$ does not depend on $T$, the value of $A$ is
twice the value of corresponding quantum correction to a single instanton
density. The calculation of two loop correction even for single instanton is
a complicated problem (see \cite{Shuryak}), but one can avoid these
calculations. The splitting of the ground state energy $\Delta E$
(\ref{eq:DE}) just equals to the twice instanton density. The semiclassical
method of finding the exponentially small splitting was given in
\cite{Landau}. The accuracy of semiclassical calculation allows to search
for any of $\sim g^{2n}$ corrections to small $\Delta E$. The first
term was found in \cite{Zinn-Justin}:
\bq\label{eq:AAA}
 E_1^1=\fr{A}{2}=-\fr{71}{12} \,\, .
 \ee

\subsection{The choice of classical trajectory}
\label{sec:1.3}

In order to introduce the collective variable describing the relative
positions of instanton and antiinstanton we multiply the functional
integral by the Faddeev-Popov unit:
\bqa\label{eq:Fadd}
1 = \int \delta(\big<\psi|x\big>-\tau)
\fr{\partial(\big<\psi|x\big> -\tau)}
{\partial T}dT \,\, , \,\,
\big< \psi|x\big> \equiv \int \psi(t)x(t)dt \,\, .
\eea
Here $T$ is the distance between pseudoparticles, $\tau = \tau(T)$ -- some
function and $\psi= \psi(t,T)$ -- a vector in the functional space. The
functional integration is to be performed over the hyperplanes orthogonal to
the vector $\psi$. We would call the instanton -- antiinstanton pair the
configuration in the functional space which minimizes the action for a given
value of collective variable $\tau(T)$. The equation of motion now reads:
\bq\label{eq:eqm}
\fr{\delta S[x_{I-A}]}{\delta x}=\xi\psi \,\, ,
\ee
where $\xi$ is a Lagrange multiplier, which allows to satisfy the condition
$\big<\psi | x_{I-A}\big> =\tau$.

In order to describe the instanton -- antiinstanton pair one have to
introduce the second collective variable $t_0$ -- the position of the center
of the pair. But at least if $\psi(t)$ is symmetric under transformation
$t-t_0\rightarrow t_0-t$, the second collective variable may be chosen in
such a way, that it will not affect the classical configuration.

There is a great freedom in the choice of $\psi(t,T)$ and $\tau(T)$
(\ref{eq:Fadd}). The only restriction is that after we have separated the
collective variables the functional integral over all the rest (quantum)
variables must be well convergent. Technically the functional integration
reduces to calculation of $det \fr{\delta^2 S[x_{I-A}]}{\delta x^2}$ -- the
determinant of the second action variation in the field of the pair. Among
the directions in the functional space there are two the "worse". These are
the almost zero modes --- the eigenfunctions of the operator $\fr{\delta^2
S}{\delta x^2}$ having anomalously small eigenvalues. The vector $\psi$ from
(\ref{eq:Fadd}) should not be orthogonal to these modes. Under this
conditions any function $\psi (t)$ would lead to a reliable instanton --
antiinstanton type solution of the equation (\ref{eq:eqm}). In the following
we suppose that $\psi (t)$ is a symmetric function, which differs from zero
only near the centers of separate instanton and antiinstanton and goes to zero
very fast both inside and outside the pair.

\subsection{ Corrections to the classical trajectory and action}
\label{sec:a}

Thus over a wide range between pseudoparticles $\psi\simeq 0 $ and eq.
(\ref{eq:eqm}) takes the form:
\bq\label{eq:yr}
-\ddot {x}_{I-A}+x_{I-A}=3gx_{I-A}^2-2g^2x_{I-A}^3 \,\, ,
\ee
Here 'dot' means the time derivative. Let us represent $x_{I-A}$ as a sum of
instanton and antiinstanton field plus some correction:
\bqa\label{eq:dxd}
 x_{I-A}&=&x_{I}(t+T/2)+x_{A}(t-T/2)+\delta x \, .
\eea
Here $x_{I}$ is an unperturbed instanton (\ref{eq:Inst}) placed at $t=-T/2$
and $x_{A}$ is
an antiinstanton placed at $t=T/2$. Solving eq. (\ref{eq:yr}) iteratively in
the intermediate region between instanton and antiinstanton, we get:
\bq\label{eq:xxx}
x_{I-A}=\fr{e^{-T/2}}{g}(e^{t}+e^{-t})+
\fr{e^{-T}}{g}(e^{2t}+e^{-2t}+6)+...\,\, , \,\,
\delta x\simeq 6e^{-T}/g\,\,.
\ee
Outside the instanton -- antiinstanton pair, at $|t|-T/2 \gg 1$, $\delta x$
is neglegibly small.

The action accurate to terms $\sim Te^{-2T}/g^2$ reads:
\bqa\label{eq:S1}
S[x_{I-A}]&\simeq& S[x_I]+S[x_A]+S[\delta x]+   \\
+ 2\biggl(\int_{-\infty}^0 dt \fr{\delta S[x_I]}{\delta x}(x_A+\delta x)
&+&\int_{-\infty}^0 dt \fr{\delta S[x_A]}{\delta x}\delta x\!
+\!(\dot{x}_I+\dot{x}_A)\delta x|_{-\infty}^0\,
+\, \dot{x}_I x_A|_{-\infty}^0+ \non \\
+\int_{-\infty}^0dt(3g^2x_I^2x_A^2-3gx_A^2x_I
 &-&6gx_Ix_A\delta x)\biggr)= \non \\
&=&\fr{1}{g^2}(1/3-2e^{-T}-12Te^{-2T}+O(e^{-2T}))\,\,. \non
\eea
Here we use $(\dot{x}_I+\dot{x}_A)\delta x|_0\equiv 0$ and the equation of
motion $\fr{\delta S[x_I]}{\delta x} \equiv 0$.
Corrections to (\ref{eq:S1}) are of the order of $\delta S\sim e^{-2T}/g^2 $
and depend on the choice of $\psi$ (\ref{eq:Fadd}),(\ref{eq:eqm}). But the
contribution $\sim Te^{-2T}/g^2$ in the formula (\ref{eq:S1}) is model
independent.

\subsection{ Corrections to determinant}
\label{sec:¡}

Now we have to find the determinant of the operator
\bq\label{eq:Oper}
M=\fr{\partial^2 S[x_{I-A}]}{\partial x^2}=-\fr{d^2}{dt^2}+
1+U_I+U_A+\Delta U \,\, , \,\,
U_{I,A}(t)=-\fr{3}{2ch(t \pm T/2)}\,\, .
\ee
The correction
$\Delta U$ equals to
\bq\label{eq:DeltaU}
 \Delta U =-24e^{-T}
\ee
over the wide range between instanton and antiinstanton ($T/2 -|t|\gg 1$)
and goes to zero outside the pair ($|t|-T/2 \gg 1$). Let us use the equality
\bq\label{eq:Det1}
det\biggl(\fr{M}{M_0}\biggr)=\exp\biggl(tr\, ln \biggl(1+\fr{U_I+U_A+\Delta U}
{-d_{tt}^2+1}\biggr)\biggr) \,\,,
\ee
where $M_0=-d_{tt}^2+1.$ Consider the logarithm in (\ref{eq:Det1}) as the
power series in the potential $U$. Because of the Green function of
the operator $M_0$ : $G_0(t,t')=e^{-|t-t'|}/2$ falls down exponentially at
large $|t-t'|$, the interference of $U_1$ and $U_2$ in (\ref{eq:Det1}) (as
well as $\Delta U$) may be neglected in the zeroth approximation. In this
approximation the total determinant factorizes to the product of the single
instanton determinants. There are two sources of the corrections to
(\ref{eq:Det1}) of the order of $Te^{-T}$. These are the contribution of the
first order in $\Delta U$ and the interference in the lowest order of $U_1$
and $U_2$:
\begin{eqnarray}\label{eq:1M}
-\int dt G(t,t)\Delta U &=&-12Te^{-T}  \,\, , \\
-\fr{1}{2} \int dt dt'G^2(t,t')(U_I(t)U_A(t') &+& U_I(t')U_A(t))=
-12Te^{-T} \,\, . \nonumber
\end{eqnarray}
The full determinant up to model dependent corrections $\sim O(e^{-T})$
reads
\bq\label{eq:DET2}
det\biggl(\fr{M}{M_0}\biggr) = det^2\biggl(\fr{-d_{tt}^2+1+U_I}
{-d_{tt}^2+1}\biggr)
(1-24Te^{-T}+O(e^{-T})) \,\, .
\ee

\section{Results}
\label{sec:4}

All the corrections (\ref{eq:rhoIA}),(\ref{eq:S1}),(\ref{eq:DET2}) should be
combined in expression like (\ref{eq:En1}) or (\ref{eq:En}). Now if one pass
again to the variable $x=g^2S(T)=1/3-2e^{-T}-12Te^{-2T}+\cdots\,\,$ the
Borel function $F(x)$ (\ref{eq:Fbor}) may be found in a form:
\bqa\label{eq:fun}
F(x) \simeq \biggl(\fr{-1}{1/3-x}+3ln(1/3-x)+\ldots\biggr)
\biggl(1-\fr{71}{6}g^2\biggr) \,\,.
\eea
The straightforward calculation leads now to the asymptotics (\ref{eq:Asip}).
The first attempt to calculate numerically the $1/n$ correction to the
asymptotics was made in \cite{Brezin}. The authors of \cite{Brezin} were the
first who have calculated the leading ($\sim n!$) asymptotics
(\ref{eq:Asip}) of the perturbative expansion of ground state energy of the
double well oscillator (\ref{eq:Ham}). However their numerical value for the
$1/n$ correction was incorrect, being $103/36$ instead of $53/18$. In Table
I. about $10$ first terms of approximate expansion (\ref{eq:Asip}) are
compared with the exact values \cite{Brezin}. Starting from $n=5$ the error
in $E_n$ never exceeds $12\%$.

It is convenient to integrate by part the $\sim \ln(1/3-x)$ term in $F(x)$
(\ref{eq:fun}). Now the expression for the "center of band" transforms to
(compare with (\ref{eq:Etoch})):
\bq\label{eq:Etoch1}
E_{cb}=\sum_{n=0}^Ng^{2n}E_n
-\biggl\{ \int_{0}^{1/3}
\fr{e^{-x/g^2}-e^{-1/3g^2}}{1/3-x}\fr{dx}{\pi g^2}
\!\,-\fr{e^{-1/3g^2}ln(6)}{\pi g^2} - \! \sum_{n=0}^N
\fr{3^{n+1} n! }{\pi} g^{2n} \biggr\} \biggl(1-\fr{53}{6}g^2\biggr)
\ee
The accuracy of this formula depends on $N$ - the number of the terms of
perturbative expansion, which were found exactly. In order to clarify this
issue let us rewrite the integral in (\ref{eq:Etoch1}) through the
principal value integral:
$$\int_{0}^{1/3}dx\fr{e^{-x/g^2}-e^{-1/3g^2}}{1/3-x} \!=\!P\int_{0}^{\infty}dx
\fr{e^{-x/g^2}}{1/3-x}\,+\,e^{-1/3g^2}(ln(3g^2)-\gamma)\,\,.$$
Now it is easy to subtract explicitly the $N$ terms of the approximate
expansion:
\bq\label{eq:Form}
P\int_{0}^{\infty} \fr{e^{-x/g^2}}{1/3-x}\fr{dx}{\pi g^2}\,-\sum_{n=0}^{N}
\fr{3^{n+1} n!}{\pi}g^{2n}
\equiv  P\int_{0}^{\infty} \fr{(3x)^Ne^{-x/3g^2}}{1/3-x} \fr{dx}{\pi g^2}
\,\,\, .
\ee
Formula (\ref{eq:Etoch1}) reaches the best accuracy when one minimizes the
value of (\ref{eq:Form}). This minimum occurs at $|N-1/3g^2|\sim 1$. For
such $N$ the integral in r.h.s. of (\ref{eq:Form}) (as well as the minimum
term of the perturbative expansion) is of the order of $\sim 1/g
e^{-1/3g^2}$. The error in $E_{cb}$ (\ref{eq:Etoch1}) is much smaller than
the minimum term of the series $\delta E_{cb} \sim g^0 e^{-1/3g^2}$. The
integral in r.h.s. of (\ref{eq:Form}) is extremely easy to estimate if
$N\equiv 1/3g^2$ with some (large) value $N$. For such $g$ the "best value"
of $E_{cb}$ reads:
\bq\label{eq:E_pol}
E_{cb}\approx\sum_{n=0}^Ng^{2n}E_n+\fr{3Ne^{-N}}{\pi}\biggl\{ln(6N)+\gamma+
\fr{1}{3}\sqrt{\fr{2\pi}{N}}\biggr\}\biggl(1-\fr{53}{18N}\biggr)
\ee
Except for the new factor $1-53/18N$ (or $1-53g^2/6$) our results
(\ref{eq:E_pol}), (\ref{eq:Etoch1}) agree with that presented in the
literature (see \cite{Zinn-Justin N.P.}). In Table II we compare the exact
(numerical) values of $E_{cb}$ with that found from the approximate formula
(\ref{eq:E_pol}) for various values of $N\equiv 1/3g^2$. It is seen that the
error in (\ref{eq:E_pol}) is of the order of $\sim e^{-N}\equiv e^{-1/3g^2}$.

\vspace{0.5cm}

{\bf Asknowledgements} Authors are grateful to V.L.Chernyak for helpful
discussions. The work of P.G.S. was supported by International Science
Foundation.

\section{Appendix}
\label{sec:5}

In this Appendix we would like to consider the three instanton contribution
to the asymptotics of $\sim g^{2n}$ corrections to the nonperturbative
splitting $\Delta E=E_--E_+$ of the Hamiltonian (\ref{eq:Ham}) ground state
energy. Previously this asymptotics was found within the usual quantum
mechanical approach in \cite{Zinn-Justin Mat}, and via analytical
continuation of the multiinstanton contribution over $g$ in
\cite{Zinn-Justin}. Although the trick with analytical continuation
\cite{Bogomol} leads to the correct asymptotics, the author of
\cite{Zinn-Justin} have not
shown, what the specific three instanton effects contribute to $\Delta E$.

Following \cite{Zinn-Justin} consider the functional integral with inverted
($x \rightarrow 1/g -x$) boundaries:
\bq\label{eq:trP}
e^{-E_+L}-e^{-E_-L}=TrPe^{-HL}=N\int_{x(-\fr{L}{2})+x(\fr{L}{2})=1/g}
Dx(t) e^{-S}\,\,\,\,\,\,\,\, ;\,\,\,\,\, L \gg 1 \,\,.
\ee
Only the $n$-instanton configurations with odd $n$ contribute to
$TrPe^{-HL}$. The single instanton contribution is trivial:
\bq\label{eq:1-ins}
\biggl(TrPe^{-HL}\biggr)^{(1-ins)}=(2e^{-L/2})L\fr{e^{-1/6g^2}}{\sqrt{\pi g}}.
\ee
There are two distinct three-instanton contributions linear in $L$. First is
the contribution from compact molecule -- like three-instanton object having
two interior degrees of freedom. The second contribution appears due to the
effect of "excluded volume". Consider the pair of instanton and
antiinstanton separated by interval $T$ from each other and the single
instanton outside the pair. While the integral over positions of the pair
gives the factor $L$, the instanton can "run" only over the volume $L-T$.
Thus the interference $\sim LT$ should be considered in addition to the
factorized $\sim L^2$ contribution of the pair and single instanton.

Like it was done for the instanton pair (\ref{eq:En1}) we have to subtract
the factorized three-instanton and instanton plus pair contributions from the
contribution of three-instanton molecule . After such regularization the
molecule gives
\bq\label{eq:I11}
I_1=(2e^{-L/2})\fr{1}{(\pi g)^{3/2}}Le^{-1/6g^2}\int\int dtdt'\biggl(
e^{-S(t,t')}-e^{-S_p(t)}-e^{-S_p(t')}+e^{-1/3g^2}\biggr)\,\, ,
\ee
where $t,t'>0$ -- are the distances between the first and second and second
and third pseudoparticles. $S(t,t')=1/g^2(1/3-2e^{-t}-2e^{-t'}+\cdots)$ is
the three-instanton action (with $1/6g^2$ excluded) and $S_p(t)=1/g^2(1/3-
2e^{-t}+\cdots) $ is the action of instanton -- antiinstanton pair
(\ref{eq:SInst}). Let us restrict the integration in (\ref{eq:I11}) by
condition $t,t'< T_{reg}$, where intermediate scale $1\ll T_{reg}\ll
e^{1/6g2}$ should cancel in the final result. Consider the integral
\bq\label{eq:In0}
K=\int \,\int dtdt'\biggl(e^{-S(t,t')}-e^{-1/3g^2}\biggr)\,\, .
\ee
Here the integration should be performed over the range $S(t,t')>0$. After
passing to variables $t$ and $x=g^2S(t,t')$ the integration over $t$ may be
done explicitly
\bq\label{eq:In2}
K=\int_0^{1/3}dx\fr{e^{-x/g^2}-e^{-1/3g^2}}{1/3-x}2\biggl(ln\biggl(\fr{1/3-x}
{2}\biggr)+T_{reg}\biggr)\,\,.
\ee
All the rest contributions in (\ref{eq:I11}) are proportional to $T_{reg}$
and naturally lead to cancellation of the proportional to $T_{reg}$ part of
(\ref{eq:In2}).

The effect of excluded volume leads to the linear in $L$ three-instanton
contribution of the form
\bq\label{eq:I2}
I_2=(2e^{-L/2})\fr{1}{(\pi g)^{3/2}}Le^{-1/6g^2}\int_{S_p(T)=0}^{\infty}
dT(-T)\biggl(e^{-S_p(T)}-e^{-1/3g^2}\biggr)\,\,.
\ee
After the change of variables $T\rightarrow x=g^2S_p(T)$ the integral in
(\ref{eq:I2}) transforms to that of (\ref{eq:In2}).

Finally the splitting of the ground state energy reads:
\bq\label{eq:DeltaE}
\Delta E=\fr{2}{(\pi g)^{3/2}}e^{-1/6g^2}
3\int_0^{1/3}dx\fr{e^{-x/g^2}-e^{-1/3g^2}}{1/3-x}ln(\fr{1/3-x}{2})
\ee
Expanding $ln(1/3-x)/(1/3-x)$ in powers of $x$ one finds the series
(\ref{eq:DE}).

\newpage

\begin{center}
\begin{tabular}{|c|c|c|c|}
\hline
  &       &        &                          \\
n & $E_n$ & $3^nn!\fr{3}{\pi}\fr{1}{E_n}$ &
 $3^nn!\fr{3}{\pi}(1-\fr{53}{18n})\fr{1}{E_n}$    \\
  &       &        &                          \\ \hline
3 & 44.5  & 3.47   & 0.06                     \\ \hline
4 & 626.6 & 2.96   & 0.78                     \\ \hline
5 & $ 1.1031\!\,\, 10^4$     & 2.52 & 1.04    \\ \hline
6 & $ 2.2888\!\,\, 10^5$     & 2.19 & 1.11    \\ \hline
7 & $ 5.4198\!\,\, 10^6$     & 1.94 & 1.12    \\ \hline
8 & $ 1.4360\!\,\, 10^8$     & 1.76 & 1.11    \\ \hline
9 & $ 4.2015\!\,\, 10^9$     & 1.62 & 1.09    \\ \hline
10 & $ 1.3448\!\,\, 10^{11}$ & 1.52 & 1.07    \\ \hline
11 &  $4.6755\!\,\, 10^{12}$ & 1.44 & 1.06    \\ \hline
\end{tabular}
\end{center}
\begin{center}
\vspace{0.2cm}
Table I
\end{center}
\vspace{0.3cm}
The exact values of the coefficients of perturbative expansion $E_n$ (see
\cite{Brezin}) and the ratio of approximate to exact value found in the
leading ($\sim n!$) approximation and with $\sim 1/n$ correction taken into
account. It is seen that starting from $n=5$ the accuracy of the approximate
formula (\ref{eq:Asip}) is not worse than 12\% .

\vspace{0.5cm}
\begin{center}
\begin{tabular}{|c|c|c|c|c|c|}
\hline
$N \equiv 1/(3g^2)$ & 4 & 6 & 8 & 10 & 12 \\ \hline
$E_{cb}^{exact}$ & 0.4439 & 0.43797 & 0.44832 & 0.459178 & 0.467156 \\ \hline
$E_{cb}^{theor}$ & 0.4367 & 0.44367 & 0.44933 & 0.459307 & 0.467173 \\ \hline
\end{tabular}
\end{center}
\begin{center}
\vspace{0.2cm}
Table II
\end{center}

\vspace{0.3cm}
The exact numerical energy $E_{cb}^{exact}$ and the approximate value
$E_{cb}^{theor}$ found from the formula (\ref{eq:E_pol}) as a function of
$g^2$. As we expect the accuracy is $ \sim e^{-N}$.

\end{document}